\documentclass[a4paper,12pt]{article}
\usepackage[utf8]{inputenc}
\usepackage[T1,T2A]{fontenc}
\usepackage{amsmath,amsfonts,amssymb}
\usepackage{cite}

\newcommand{\lb}[0]{\left(}
\newcommand{\rb}[0]{\right)}
\newcommand{\lsb}[0]{\left[}
\newcommand{\rsb}[0]{\right]}

\newcommand{\veps}{\varepsilon}
\newcommand{\pz}{\partial_z}
\newcommand{\pzz}{\partial^2_{zz}}

\begin{document}

\renewcommand*{\thefootnote}{\fnsymbol{footnote}}

\begin{center}
{\large\bf  Confinement Potential in a Soft-Wall Holographic Model with a Hydrogen-like Spectrum}
\end{center}
\begin{center}
{Sergey Afonin\footnote{E-mail: \texttt{s.afonin@spbu.ru}} and
Timofey Solomko
}
\end{center}

\renewcommand*{\thefootnote}{\arabic{footnote}}
\setcounter{footnote}{0}

\begin{center}
{\small Saint Petersburg State University, 7/9 Universitetskaya nab.,
St.Petersburg, 199034, Russia}
\end{center}

\bigskip

\begin{abstract}
It is well known that the Soft-Wall holographic model for QCD successfully reproduces not only
the linear Regge spectrum, but also, via the holographic Wilson confinement criterion,
the ``linear plus Coulomb'' confinement potential similar to the Cornell potential. This property could be
interpreted as a holographic counterpart of the hadron string picture, where the linearly
rising potential and Regge-like spectrum are directly related. However, such a relation
does not take place in the bottom-up holographic approach. Namely, the Cornell-like potentials arise in a broad
class of bottom-up holographic models, the standard Soft-Wall model is merely a particular representative
of this class. 
We consider a calculation of confinement potential within a different representative
of this class --- a Soft-Wall model with linear dilaton background in the metric. This model leads
to a Hydrogen like spectrum.
The calculation of renormalized potential at short distances turns out to be complicated by a new subtlety that
was skipped in general discussions of the issue existing in the literature.
But the confinement potential of the model is shown to be not very
different from the potential obtained in the standard Soft-Wall model with quadratic background.
\end{abstract}


\section{Introduction}

The Soft-Wall (SW) holographic model has enjoyed a considerable phenomenological success
in applications to various problems in the physics of strong interactions (see, e.g.,
a review of recent literature in~\cite{Afonin:2021cwo}). The SW model was originally introduced
in~\cite{son2} and~\cite{andreev} as a bottom-up holographic model with a certain static
gravitational background that reproduces the Regge spectrum of light mesons.

This background does not represent a solution of some known full-fledged dual 5D gravitational theory,
rather it interpolates in a model way some important contributions from a hypothetical unknown dual
string theory for QCD. The crux of the matter is that QCD is weakly coupled at high energy due to
asymptotic freedom and the dual description of weakly coupled theories requires highly curved space-times
for which the approximation by usual general relativity is not enough to analyze the dual theory.
As their dual formulations cannot be given in terms of a gravitational field theory one must involve,
strictly speaking, the full dual string theory~\cite{PhysToday}. This task is too difficult to implement
but phenomenologically one can guess a background that correctly reproduces the high energy behavior of
two-point correlation functions in QCD together with leading power-like non-perturbative corrections
known from QCD sum rules.
Such a background was invented in~\cite{son2,andreev} that gave rise to the SW holographic approach.

Among numerous phenomenological applications of SW model which rapid\-ly followed after its inception
there was a derivation of heavy-quark potential by Andreev and Zakharov~\cite{Andreev:2006ct}
which will be in the focus of our present work. The found potential turned out to be similar to the
well-known Cornell potential~\cite{bali}. The paper~\cite{Andreev:2006ct} was followed by many works
in which the derivation of confinement potential was analyzed within various AdS/QCD
models~\cite{heavy1,heavy2,heavy3,heavy4,heavy5,heavy6,heavy7,heavy8,Afonin:2021zdu,Afonin:2022aqt}.
They are based on a general result proved in Ref.~\cite{Kinar:1998vq} that Cornell-like behavior
arises from the AdS/CFT correspondence using quite general conditions on the geometry which
were found. A good review of the issue is given in Ref.~\cite{heavy5}.
The potential derived in~\cite{Andreev:2006ct} was quantitatively compared with
potentials obtained within some general AdS deformed metrics in~\cite{heavy2}
and it was found that a geometry based on a simple quadratic warp factor used in~\cite{Andreev:2006ct}
formally agrees most closely with the data in comparison with more complicated geometries.

In the given paper, we discuss the holographic relationship between the linear confinement
in the sense of area law for Wilson loops and the Regge behavior of mass spectrum. They turn out to
be completely unrelated in the bottom-up holographic approach: One can construct an infinite
number of bottom-up holographic models which do not lead to Regge spectrum but do have a property
of linear confinement. As an example, we analyze in detail a SW holographic model with linear
exponential background in the AdS metric. The spectrum of this model is Hydrogen like, nevertheless
we show by an explicit calculation that the heavy-quark potential following from the model
is close to the potential derived within the standard SW model with quadratic exponential background
in~\cite{Andreev:2006ct} and, for a more general case, in~\cite{Afonin:2021zdu,Afonin:2022aqt}.
Our main results have been briefly announced in the conference paper~\cite{qfthepc}, in the given
work we expand the discussions and provide detailed calculations.

The paper is organized as follows. The simplest scalar SW holographic models with quadratic and linear
exponential background in the AdS metric are reminded in Section~2. In Section~3, we briefly review the
derivation of confinement potential from the holographic Wilson loop and discuss some important consequences.
Then, in Section~4, we show a detailed calculation of confinement potential in the scalar SW model with
linear exponential background. The summary and some concluding remarks are provided in the final Section~5.

\section{The simplest SW holographic models with quadratic and linear exponential background}

Perhaps the simplest variant of the Soft-Wall holographic model is given by the following 5D
action for a real massless scalar field $\Phi$,
\begin{equation}
\label{sw1}
  S=\int d^4\!x\,dz\sqrt{g}\,g^{MN}\partial_M\Phi\partial_N\Phi,
\end{equation}
where $g=|\text{det}g_{MN}|$, the metric is given by a modification of the Poincar\'{e} patch of the AdS$_5$ space,
\begin{equation}
\label{az_metric}
ds^2=g_{MN}dx^M\!dx^N=h\frac{R^2}{z^2}\lb\eta_{\mu\nu}dx^\mu\! dx^\nu-dz^2\rb,
\end{equation}
\begin{equation}
\label{h}
  h=e^{2cz^2/3}.
\end{equation}
Here $\eta_{\mu\nu}=\text{diag}\lbrace1,-1,-1,-1\rbrace$, $z>0$ is the holographic coordinate,
$R$ denotes the radius of AdS$_5$ space, and $c>0$ represents
a parameter of dimension of mass squared that introduces the mass scale into the model.
The factor of $2/3$ in~\eqref{h} is chosen for our convenience. The SW holographic model in this form was first
proposed in~\cite{andreev} (a massless vector field and $c<0$ were used in the formulation of Ref.~\cite{andreev}).

The corresponding equation of motion,
\begin{equation}
  \partial_N\lb\sqrt{g}\,g^{MN}\!\partial_M\Phi\rb=0,
\end{equation}
takes the form,
\begin{equation}
\label{eom0}
  \partial^\mu\partial_\mu\Phi-z^3e^{-cz^2}\pz\lb\frac{e^{cz^2}}{z^3}\pz\Phi\rb=0.
\end{equation}
It is evident that the same equation of motion follows from the action
\begin{equation}
\label{sw2}
  S=\int d^4\!x\,dz\sqrt{g}\,e^{cz^2}\!g^{MN}\partial_M\Phi\partial_N\Phi,
\end{equation}
in which the metric modification~\eqref{h} is replaced by the ``dilaton'' background $e^{cz^2}$ in the action.
This formulation of SW holographic model was put forward in~\cite{son2} for massless vector fields and
$c<0$. Its extension to the massless scalar fields (as in~\eqref{sw2} but again with $c<0$) was first
studied in~\cite{Colangelo}.
Now the formulations with dilaton background of the kind~\eqref{sw2} are most frequently used in the literature,
with various Lagrangians, backgrounds, and different signs of the mass parameter $c$.
Redefining the field \(\Phi=z^{3/2}e^{-cz^2/2}\phi\) in~\eqref{eom0} we get
\begin{equation}
  \partial^\mu\partial_\mu\phi-\pzz\phi+
  \lb c^2z^2+\frac{15}{4z^2}-2c\rb\phi=0
\end{equation}
The free hadronic states in holographic QCD are described by the plane-wave ansatz in physical
space-time,
\begin{equation}
\phi(x,z)=e^{iqx}\psi(z),\qquad q^2=m^2,
\end{equation}
where $\psi(z)$ is a $z$-dependent profile function.
This ansatz leads to the one-dimensional Schr\"{o}dinger equation for the mass squared,
\begin{equation}
\label{eom}
  -\psi''+V(z)\psi=m^2\psi,
\end{equation}
with the potential of the harmonic oscillator type,
\begin{equation}
\label{eom_pot}
  V(z)=c^2z^2+\frac{15}{4z^2}-2c.
\end{equation}
The corresponding normalizable solutions yield the discrete Regge-like mass spectrum,
\begin{equation}
\label{spSW}
m_n^2=4c(n+1),\qquad n=0,1,2,\dots.
\end{equation}

We define the Soft-Wall holographic model with linear dilaton via the replacement of function $h(z)$
in~\eqref{h} by
\begin{equation}
\label{h2}
  h=e^{4cz/3}.
\end{equation}
The factor of $4/3$ in~\eqref{h2} is again chosen for simplification of the expressions which will follow.
The mass parameter $c>0$ has now the dimension of linear mass. Equivalently, the model can be defined
via the replacement of quadratic dilaton background $e^{cz^2}$ in the action~\eqref{sw2} by the linear
dilaton background $e^{2cz}$. Repeating the same steps as above we arrive (making use of the field
redefinition \(\Phi=z^{3/2}e^{-cz}\phi\)) at the Schr\"{o}dinger equation~\eqref{eom}
with the potential
\begin{equation}
\label{eom_pot2}
V(z)=\frac{15}{4z^2}-\frac{3c}{z}+c^2.
\end{equation}
This potential is identical to the potential for the radial wave function in the Coulomb problem.
The corresponding discrete spectrum is
\begin{equation}
  m_n^2=c^2-\frac{9c^2}{4(n+k)^2},\qquad n=0,1,2,\dots,
\end{equation}
where \(k\) is the positive solution to the condition on the existence of normalizable
solutions\footnote{In the Coulomb problem, this condition is \(k(k-1)=l(l+1)\), where $l$ is the orbital quantum number,
thus $k=l+1$ and one obtains the famous Coulomb degeneracy between the radial and orbital excitations.},
\begin{equation}
k(k-1)=15/4,
\end{equation}
namely \(k=5/2\).

\section{The confinement potential from holographic Wilson loop}

The holographic derivation of a static potential between two heavy sources was
originally proposed by Maldacena in~\cite{Maldacena:1998im}. In brief, one considers a Wilson loop \(W\) situated on
the 4D boundary of Euclidean 5D space with the Euclidean time coordinate \(0\leq t \leq T\) and
the remaining 3D spatial coordinates \(-r/2 \leq y \leq r/2\). Such Wilson loop can be related to the propagation
of a massive quark~\cite{Maldacena:1998im}. The expectation value
of the loop in the limit of \(T\to\infty\) is equal to \(\left\langle W\right\rangle\sim e^{-TE(r)}\),
where \(E(r)\) is interpreted as the energy of the quark-antiquark pair. This
expectation value can be also obtained via \(\left\langle W\right\rangle\sim e^{-S}\),
where \(S\) represents the area of a string world-sheet which produces the loop
\(W\). Combining these two expressions one gets the energy of the configuration, \(E=S/T\).
A natural choice for the world-sheet area is the Nambu-Goto
action,
\begin{equation}
\label{ng}
  S=\frac{1}{2\pi\alpha'}\int d^2\xi\sqrt{\det \left[G_{MN}\partial_\alpha X^M\!\partial_\beta X^N\right]},
\end{equation}
where \(\alpha'\) is the inverse string tension, \(X^M\) are the string coordinates
functions which  map the parameter space of the
world-sheet \(\lb\xi_1,\xi_2\rb\)  into the space-time, and \(G_{MN}\) is the Euclidean
metric of the bulk space.

This idea was applied to the vector SW holographic model in~\cite{Andreev:2006ct}.
The asymptotics of the obtained potential at large and small distances qualitatively
reproduced the Cornell potential,
\begin{equation}
\label{cornell}
V(r)=-\frac{\kappa}{r}+\sigma r + \text{const},
\end{equation}
which was accurately measured in the lattice simulations and is widely used in the heavy-meson spectroscopy~\cite{bali}.
The calculation of Ref.~\cite{Andreev:2006ct} was further extended to the case of
SW models generalized to arbitrary intercept parameter (i.e., when the linear Regge spectrum
has a general form $m^2_n\sim an+b$, where the intercept $b$ is arbitrary) and to the scalar SW model
in~\cite{Afonin:2021zdu,Afonin:2022aqt}, where many details
of the derivation can be found. Below we briefly summarize the results.

The Euclidean version of the metric~\eqref{az_metric} takes the form
\begin{equation}
\label{metric}
  G_{MN}=\text{diag}\left\lbrace\frac{R^2}{z^2}h,\dots,\frac{R^2}{z^2}h\right\rbrace.
\end{equation}
The string world-sheet~\eqref{ng} has a well-known property of  reparameterization invariance.
Choosing the parametrization \(\xi_1=t\) and \(\xi_2=y\) and integrating
over \(t\) from \(0\) to \(T\) in~\eqref{ng}, one arrives at the action
\begin{equation}\label{ng_action}
  S=\frac{TR^2}{2\pi\alpha'}\int\displaylimits_{-r/2}^{r/2}dy\,
  \frac{h}{z^2}\sqrt{1+z'^2},
\end{equation}
where \(z'=dz/dy\). The action is translationally invariant (there is no explicit dependence
of the Lagrangian on \(y\)), hence there exists a conserving quantity which represents the first integral
of equation of motion,
\begin{equation}
\label{cons}
  \frac{h}{z^2}\frac{1}{\sqrt{1+z'^2}}=\text{Const}.
\end{equation}
Since \(z=0\) at the ends of the Wilson loop, \(y=\pm r/2\), and the system
is symmetric, $z$ has the maximum value at $y=0$,
\begin{equation}
\label{not_0}
  z_0\equiv\left.z\right|_{y=0}.
\end{equation}
The integration constant in~\eqref{cons} can be expressed via \(z_0\). Using \(\left.z'\right|_{x=0}=0\)
in~\eqref{cons} one can obtain the following expression for the distance \(r\),
\begin{equation}
\label{r_int_def}
  r=2\sqrt{\frac{\lambda}{c}}\int\displaylimits_{0}^{1}dv\,\frac{h_0}{h}
  \frac{v^2}{\sqrt{1-v^4\frac{h_0^2}{h^2}}},
\end{equation}
where
\begin{equation}
\label{not}
  h_0=\left.h\right|_{z=z_0},\quad
  v= z/z_0,\quad
  \lambda= cz_0^2.
\end{equation}

The expression for the energy can be derived from \(E=S/T\) using the
action~\eqref{ng_action} and expression~\eqref{r_int_def}, the relevant
details can be found, e.g., in~\cite{Afonin:2021zdu}. The final result is
\begin{equation}
\label{E2}
  E=\frac{R^2}{\pi\alpha'}\sqrt{\frac{c}{\lambda}}\int\displaylimits_{0}^{1}\frac{dv}{v^2}\,
  \frac{h}{\sqrt{1-v^4\frac{h_0^2}{h^2}}}.
\end{equation}
The integral in~\eqref{E2} is divergent at \(v=0\) but it can be regularized by imposing a cutoff
\(\veps\to 0\),
\begin{equation}
  E=\frac{R^2}{\pi\alpha'}\sqrt{\frac{c}{\lambda}}\int\displaylimits_{0}^{1}
  \frac{dv}{v^2}\lb\frac{h(\lambda,v)}{\sqrt{1-v^4\frac{h_0^2}{h^2}}}-D\rb+
  \frac{R^2}{\pi\alpha'}\sqrt{\frac{c}{\lambda}}\,D\!
  \int\displaylimits_{\veps/z_0}^{1}\frac{dv}{v^2},
\end{equation}
and introducing the regularized energy $E_R$,
\begin{equation}
\label{ER}
  E_R=\frac{R^2D}{\pi\alpha'\veps}+E,\qquad D\equiv \left.h\right|_{v=0},
\end{equation}
where $D$ is the regularization constant. One can argue that the infinite constant
in $E_R$ is related with the quark mass and must be subtracted when calculating the
static interaction energy~\cite{Kinar:1998vq}.
The resulting finite energy is
\begin{equation}
\label{en_int_def}
  E=\frac{R^2}{\pi\alpha'}\sqrt{\frac{c}{\lambda}}\lsb\int\displaylimits_{0}^{1}
  \frac{dv}{v^2}\lb\frac{h}{\sqrt{1-v^4\frac{h_0^2}{h^2}}}-D\rb-D\rsb.
\end{equation}

The expressions~\eqref{r_int_def} and~\eqref{en_int_def} give the static energy $E(r)$
as a function of distance between sources.
For the quadratic dilaton correction to the AdS metric~\eqref{h} the function $E(r)$
cannot be found analytically. But one can calculate the asymptotics at large and small $r$,
the results for the simplest scalar SW model are as follows~\cite{Afonin:2022aqt},
\begin{equation}
\label{large_r_not_s}
  E\underset{r\to\infty}{=}\frac{R^2}{\alpha'}\sigma_\infty r,\qquad \sigma_\infty=\frac{ec}{3\pi},
\end{equation}
\begin{equation}
\label{small_r_not_s}
  E\underset{r\to0}{=}\frac{R^2}{\alpha'}\lsb-\frac{\kappa_0}{r}+\sigma_0 r\rsb,
\end{equation}
where
\begin{equation}
\label{rho}
  \kappa_0=\frac{1}{2\pi\rho^2},\qquad
  \sigma_0=\frac{c\rho^2}{3},\qquad \rho=\frac{\Gamma\!\!\lb\frac{1}{4}\rb^2}{(2\pi)^{3/2}}.
\end{equation}

The expressions~\eqref{r_int_def} and~\eqref{en_int_def} have a physical meaning only if
they are real-valued, i.e., if the expression under the square root is positive.
One can show~\cite{Afonin:2022aqt} that this restriction leads to the condition
\begin{equation}
\label{sonn_cond}
  \left.\pz G_{00}\right|_{z=z_0}=0,\qquad
  \left.G_{00}\right|_{z=z_0}\ne0.
\end{equation}
In fact, this condition represents a general statement first derived in~\cite{Kinar:1998vq}
(and formulated in a more clear and concise way in~\cite{Sonnenschein:2000qm}):
The \(G_{00}\) element of the background metric dual to a confining string
theory in the sense of the area law behavior of a Wilson loop must satisfy~\eqref{sonn_cond}.

The condition~\eqref{sonn_cond} can be given an heuristic physical interpretation~\cite{br3}.
The time-time component of metric is directly related with the gravitational potential energy $U$
for a body of mass $m$, $U=mc^2\sqrt{g_{00}}$. The confinement behavior can be
qualitatively deduced from following a particle in the AdS space as it goes to the infrared region (large $z$)
--- in general relativity, this would correspond to falling an object by the effects of gravity.
If the potential $U(z)$ has an absolute minimum at some $z_0$ then a particle is confined within distances $z\sim z_0$.
In this situation, one can think of a particle confined effectively in a hadron of size $z_0$. Such a physical picture
has a nice visualization within the light-front holographic approach, where the holographic coordinate $z$ is proportional
to the interquark distance in a hadron~\cite{br3}. The condition~\eqref{sonn_cond} becomes
nothing but the condition for extremum of the potential $U(z)=mc^2\sqrt{g_{00}(z)}$.

The analysis of Ref.~\cite{Kinar:1998vq} (see also a brief summary in~\cite{Sonnenschein:2000qm})
contains another general result:
The large-distance asymptotics of potential energy in a confining string theory is given by
\begin{equation}
\label{large_r_not_s2}
  E\underset{r\to\infty}{=}\frac{G_{00}(z_0)}{2\pi\alpha'}r.
\end{equation}
This asymptotics can be directly derived from~\eqref{r_int_def} and~\eqref{E2} by expanding
the corresponding integrands at \(v=1\), expressing the integral in terms of
\(r\), and then substituting it into \(E\) (this procedure will be demonstrated below
for the SW model with linear dilaton). Since $G_{00}$ in~\eqref{h} reaches its minimum value at $z_0^2=\frac{3}{2c}$,
we get $G_{00}(z_0)=2ecR^2/3$ and the relation~\eqref{large_r_not_s2} leads immediately to the
large-distance asymptotics displayed in~\eqref{large_r_not_s}.

A theorem of Ref.~\cite{Kinar:1998vq} states that~\eqref{sonn_cond} is a sufficient condition for linear
confinement in the sense of area law for Wilson loops. It is obvious that the SW holographic model with positive
quadratic dilaton represents just a particular case when the linear confinement holds. One can easily
construct an infinite number of other bottom-up holographic models with linear confinement. In particular,
the quadratic function $z^2$ in~\eqref{h} may be replaced by $z^\alpha$, $\alpha>0$, i.e., by any positive power of $z$.
If $\alpha\neq2$ then the spectrum will not have a Regge form. Thus, {\it the linear confinement in the bottom-up
holographic approach is not related with the Regge behavior of mass spectrum}.

The Coulomb behavior at small distances is governed by the AdS metric at small $z$. Therefore, if the ultraviolet
AdS asymptotics is not violated (and it cannot be violated in sensible holographic models because otherwise the holographic
dictionary from the AdS/CFT correspondence is lost) the Coulomb short-distance asymptotics will always be reproduced.

The discussion above shows that the asymptotic structure of Cornell potential~\eqref{cornell} can be reproduced in a broad class
of bottom-up holographic models. In particular, the choice~\eqref{h2} of function $h(z)$ in the metric~\eqref{metric},
corresponding to the linear dilaton, $\alpha=1$, will also result in a Cornell-like potential. Below we derive this
potential for the given special case and finally estimate a quantitative difference with the case of quadratic dilaton.

\section{Confinement potential in the scalar SW model with linear exponential background}

\subsection{Expressions for distance and energy}

Consider the scalar SW model of Section~2 with linear exponential background~\eqref{h2} in the metric~\eqref{metric}.
Introducing the notations
\begin{equation}
  \lambda = cz_0,\qquad
  v =z/z_0,\qquad
  h_0 = e^{4cz_0/3}=e^{4\lambda/3},
\end{equation}
the expressions for the distance~\eqref{r_int_def} and energy~\eqref{E2}
take the form
\begin{equation}
\label{rl}
  r=
  \frac{2\lambda}{c}\int\displaylimits_{0}^{1}dv\,\frac{v^2e^{4\lambda(1-v)/3}}{\sqrt{1-v^4e^{8\lambda(1-v)/3}}},
\end{equation}
\begin{equation}
\label{unreg_en}
  E=
  \frac{R^2}{\pi\alpha'}\frac{c}{\lambda}\int\displaylimits_{0}^{1}dv\,
  \frac{e^{4\lambda v/3}}{v^2\sqrt{1-v^4e^{8\lambda(1-v)/3}}}.
\end{equation}
A simple analysis of the reality condition for two expressions above,
\begin{equation}
\label{sqrt_exp}
  1-v^4e^{8\lambda(1-v)/3}\ge 0,\qquad
  v\in\lsb0,1\rsb,
\end{equation}
shows that the given condition is satisfied if
\begin{equation}
\label{rc}
  0\leq\lambda<\frac{3}{2}.
\end{equation}

The integral for potential energy~\eqref{unreg_en} must be regularized in order to eliminate
the divergence at \(v=0\). This should be done carefully so that all emerging divergent terms
are properly subtracted. First, we expand the integrand of~\eqref{unreg_en} at \(v\to0\)
(note that in the series expansion of the square root in the denominator its first, second,
and third order derivatives are all equal to zero at \(v=0\)),
\begin{equation}
\label{vt}
  \frac{e^{4\lambda v/3}}{v^2\sqrt{1-v^4e^{8\lambda(1-v)/3}}}\underset{v\to0}{=}
  \frac{1}{v^2}+\frac{4\lambda}{3}\frac{1}{v}+\mathcal{O}(1).
\end{equation}
From~\eqref{vt} it follows that in order to regularize the energy we must subtract two terms,
\begin{multline}
  E_R=\frac{R^2}{\pi\alpha'}\frac{c}{\lambda}\int\displaylimits_{0}^{1}dv\,
  \lb\frac{e^{4\lambda v/3}}{v^2\sqrt{1-v^4e^{8\lambda(1-v)/3}}}-\frac{1}{v^2}-
  \frac{4\lambda}{3}\frac{1}{v}\rb+\\
  +\frac{R^2}{\pi\alpha'}\frac{c}{\lambda}\int\displaylimits_\veps^1dv\lb\frac{1}{v^2}+\frac{4\lambda}{3}\frac{1}{v}\rb.
\end{multline}
The second integral here can be easily calculated,
\begin{equation}\label{eps_int}
  \frac{R^2}{\pi\alpha'}\frac{c}{\lambda}\int\displaylimits_\veps^1dv\lb\frac{1}{v^2}+\frac{4\lambda}{3}\frac{1}{v}\rb=
  \frac{R^2}{\pi\alpha'}\frac{c}{\lambda}\lb-1+\frac{1}{\veps}-\frac{4\lambda}{3}\ln\veps\rb,
\end{equation}
and we get the regularized energy,
\begin{equation}
\label{reg_en}
  E_R=\frac{R^2}{\pi\alpha'}\frac{c}{\lambda}\lb\frac{1}{\veps}-\frac{4\lambda}{3}\ln\veps\rb+E,
\end{equation}
where the physical finite energy \(E\) is
\begin{equation}
\label{en_int}
  E=\frac{R^2}{\pi\alpha'}\frac{c}{\lambda}\lsb\int\displaylimits_{0}^{1}\frac{dv}{v^2}\,
  \lb\frac{e^{4\lambda v/3}}{\sqrt{1-v^4e^{8\lambda(1-v)/3}}}-1-\frac{4\lambda}{3}v\rb-1\rsb.
\end{equation}
Note that compared to the previous case~\eqref{ER} the \(\veps\)-terms in the regularized
energy~\eqref{reg_en} depend on \(\lambda\) or, equivalently, on \(z_0\).

\subsection{Asymptotics of energy at small distances}


The small distances in the energy~\eqref{en_int} correspond to small \(\lambda\). Thus the small distance behavior
of the potential energy can be found from the asymptotic expansion at \(\lambda\to 0\) of the expressions
for distance and energy. Expanding the integrand in~\eqref{rl} at \(\lambda\to 0\) we get (we will need
terms up to \(\mathcal{O}(\lambda^2)\), the \(\mathcal{O}(\lambda^3)\) terms are neglected),
\begin{equation}
  r\underset{\lambda\to0}{=}\frac{2\lambda}{c}\int\displaylimits_{0}^{1}dv\lsb
  \frac{v^2}{\sqrt{1-v^4}}+\frac{4\lambda}{3}\frac{v^2(1-v)}{(1-v^4)^{3/2}}+
  \frac{8\lambda^2}{9}\frac{v^2(1+2v^4)(1-v)^2}{(1-v^4)^{5/2}}\rsb\!\!.
\end{equation}
The result of integration can be written in terms of the incomplete beta function as
\begin{equation}
\label{beta}
  \int_0^1dv\,v^a(1-v^4)^b=\frac{1}{4}\lim_{x\to1}B_x\lb\frac{a+1}{4},b+1\rb,
\end{equation}
which lead to the following expression,
\begin{multline}
  r\underset{\lambda\to0}{=}\frac{\lambda}{2c}\lim_{x\to1}\left\lbrace
  B_x\lb\frac{3}{4},\frac{1}{2}\rb+
  \frac{4\lambda}{3}\lsb B_x\lb\frac{3}{4},-\frac{1}{2}\rb-B_x\lb1,-\frac{1}{2}\rb\rsb+\right.\\\left.+
  \frac{8\lambda^2}{9}\lsb B_x\lb\frac{3}{4},-\frac{3}{2}\rb-2B_x\lb1,-\frac{3}{2}\rb+
  B_x\lb\frac{5}{4},-\frac{3}{2}\rb+2B_x\lb\frac{7}{4},-\frac{3}{2}\rb-\right.\right.\\\left.\left.-
  4B_x\lb2,-\frac{3}{2}\rb+2B_x\lb\frac{9}{4},-\frac{3}{2}\rb\rsb\right\rbrace.
\end{multline}
The first beta function has two positive arguments, so it does not contain any
divergences. To show that the divergences from other beta functions cancel each other out
we make use of the following expansion for the incomplete beta function,
\begin{equation}
\label{beta_s1}
  B_x(a,b)\underset{x\to1}{=}B(a,b)-\frac{(1-x)^b}{b}-\frac{(a+b)(1-x)^{b+1}}{b(b+1)}+\mathcal{O}\lb(x-1)^{b+2}\rb.
\end{equation}
We can see immediately that for the incomplete beta functions inside the first square brackets only
the first two terms in the expansion~\eqref{beta_s1} are important. In addition, since
the second arguments of these beta functions are equal and the beta functions have
opposite signs, the divergences stemming from the second term of the expansion~\eqref{beta_s1}
cancel each other out.

The expansions for each of the incomplete beta functions inside the second square brackets
are
\begin{equation}
\label{beta_exps}
  \begin{aligned}
    B_x\lb\frac{3}{4},-\frac{3}{2}\rb&\underset{x\to1}{=}B\lb\frac{3}{4},-\frac{3}{2}\rb+
    \frac{2}{3}(1-x)^{-3/2}+(1-x)^{-1/2}+\mathcal{O}\lb(x-1)^{1/2}\rb,\\
    B_x\lb1,-\frac{3}{2}\rb&\underset{x\to1}{=}B\lb1,-\frac{3}{2}\rb+
    \frac{2}{3}(1-x)^{-3/2}+\frac{2}{3}(1-x)^{-1/2}+\mathcal{O}\lb(x-1)^{1/2}\rb,\\
    B_x\lb\frac{5}{4},-\frac{3}{2}\rb&\underset{x\to1}{=}B\lb\frac{5}{4},-\frac{3}{2}\rb+
    \frac{2}{3}(1-x)^{-3/2}+\frac{1}{3}(1-x)^{-1/2}+\mathcal{O}\lb(x-1)^{1/2}\rb,\\
    B_x\lb\frac{7}{4},-\frac{3}{2}\rb&\underset{x\to1}{=}B\lb\frac{7}{4},-\frac{3}{2}\rb+
    \frac{2}{3}(1-x)^{-3/2}-\frac{1}{3}(1-x)^{-1/2}+\mathcal{O}\lb(x-1)^{1/2}\rb,\\
    B_x\lb2,-\frac{3}{2}\rb&\underset{x\to1}{=}B\lb2,-\frac{3}{2}\rb+
    \frac{2}{3}(1-x)^{-3/2}-\frac{2}{3}(1-x)^{-1/2}+\mathcal{O}\lb(x-1)^{1/2}\rb,\\
    B_x\lb\frac{9}{4},-\frac{3}{2}\rb&\underset{x\to1}{=}B\lb\frac{9}{4},-\frac{3}{2}\rb+
    \frac{2}{3}(1-x)^{-3/2}-(1-x)^{-1/2}+\mathcal{O}\lb(x-1)^{1/2}\rb.
  \end{aligned}
\end{equation}
Taking into account the coefficients in front of the beta functions we can see
the cancelation of all divergent terms, schematically
\begin{equation}
  \begin{aligned}
    1-2+1+2-4+2=0,\\
    1-2\cdot\frac{2}{3}+\frac{1}{3}+2\cdot\frac{-1}{3}-4\cdot\frac{-2}{3}-2=0.
  \end{aligned}
\end{equation}
Thus, we can switch to the usual ``complete'' beta functions.

The remaining beta functions are then equal to
\begin{equation}
\label{beta_values1}
  \begin{aligned}
    &B\lb\frac{3}{4},\frac{1}{2}\rb=\frac{2}{\rho},\quad
    B\lb\frac{3}{4},-\frac{1}{2}\rb=-\frac{1}{\rho},\quad
    B\lb1,-\frac{1}{2}\rb=-2,\\
    &B\lb\frac{3}{4},-\frac{3}{2}\rb=-\frac{1}{2\rho},\quad
    B\lb1,-\frac{3}{2}\rb=-\frac{2}{3},\quad
    B\lb\frac{5}{4},-\frac{3}{2}\rb=-\frac{\pi\rho}{6},\\
    &B\lb\frac{7}{4},-\frac{3}{2}\rb=\frac{1}{2\rho},\quad
    B\lb2,-\frac{3}{2}\rb=\frac{4}{3},\quad
    B\lb\frac{9}{4},-\frac{3}{2}\rb=\frac{5\pi\rho}{6},
  \end{aligned}
\end{equation}
where the constant $\rho$ is the same as in~\eqref{rho}.
Putting it all together, we obtain
\begin{equation}
\label{dist_small_l}
  r\underset{\lambda\to0}{=}
  \frac{1}{\rho}\frac{\lambda}{c}\lsb1+\frac{2\lambda}{3}\lb2\rho-1\rb+
  \frac{4\lambda^2}{9}\lb\frac{1}{2}-4\rho+\frac{3\pi\rho^2}{2}\rb\rsb.
\end{equation}


Now consider the ultraviolet asymptotics of the energy. Expanding the integrand of~\eqref{en_int} at
\(\lambda\to0\) up to $\mathcal{O}(\lambda^2)$ terms we get
\begin{multline}
  E\underset{\lambda\to0}{=}\frac{R^2}{\pi\alpha'}\frac{c}{\lambda}\lsb\int\displaylimits_{0}^{1}dv\,\lb
  \frac{1}{v^2}\lb\frac{1}{\sqrt{1-v^4}}-1\rb+\right.\right.\\\left.\left.+
  \frac{4\lambda}{3}\lb\frac{1}{v}\lb\frac{1}{\sqrt{1-v^4}}-1\rb+\frac{v^2-v^3}{(1-v^4)^{3/2}}\rb+\right.\right.\\\left.\left.
  \frac{8\lambda^2}{9}\frac{1+2v^2-2v^3-2v^4+v^6-4v^7+4v^8}{(1-v^4)^{5/2}}+
  \rb-1\rsb.
\label{El}
\end{multline}
The first term in~\eqref{El} was integrated before, the result is
\begin{equation}
  \int\displaylimits_{0}^{1}\frac{dv}{v^2}\lb\frac{1}{\sqrt{1-v^4}}-1\rb=
  \frac{1}{4}B\lb-\frac{1}{4},\frac{1}{2}\rb+1.
\end{equation}

The second term in~\eqref{El} can be rewritten as
\begin{multline}
  \int\displaylimits_{0}^{1}\frac{dv}{v}\lb\frac{1}{\sqrt{1-v^4}}-1\rb=
  \lim_{\substack{x\to1\\y\to0}}\lim_{\veps\to0}\lsb\int\displaylimits_{y}^{x}
  \frac{dv}{v^{1-\veps}\sqrt{1-v^4}}-\int\displaylimits_{y}^{x}\frac{dv}{v}\rsb=\\=
  \lim_{\substack{x\to1\\y\to0}}\lim_{\veps\to0}\lsb
  \int\displaylimits_{0}^{x}\frac{dv}{v^{1-\veps}\sqrt{1-v^4}}-
  \int\displaylimits_{0}^{y}\frac{dv}{v^{1-\veps}\sqrt{1-v^4}}-
  \int\displaylimits_{y}^{x}\frac{dv}{v}\rsb=\dots
\end{multline}
We can now integrate using~\eqref{beta} and we get
\begin{equation}
  \dots=\frac{1}{4}\lim_{\substack{x\to1\\y\to0}}\lim_{\veps\to0}\lsb
  B_{x}\lb\veps,\frac{1}{2}\rb-B_{y}\lb\veps,\frac{1}{2}\rb-\ln x+\ln y\rsb=\dots
\end{equation}
Using the expansions~\eqref{beta_s1} and
\begin{equation}
  B_y(a,b)\underset{y\to0}{=}\frac{y^a}{a}+\mathcal{O}(y^{a+1}),
\end{equation}
it is seen that all divergences cancel each other out,
\begin{multline}
  \dots=\frac{1}{4}\lim_{\substack{x\to1\\y\to0}}\lim_{\veps\to0}\lsb
  B\lb\veps,\frac{1}{2}\rb-2\sqrt{1-x}+\mathcal{O}\lb(x-1)^{3/2}\rb-\right.\\\left.-
  \frac{y^\veps}{\veps}+\mathcal{O}(y^{\veps+1})-\ln x+\ln y\rsb=\dots
\end{multline}
The \(x\to1\) limit can be taken safely.
Expanding the remaining terms at \(\veps\to0\) we get
\begin{equation}
  \dots=\frac{1}{4}\lim_{\substack{\veps\to0\\y\to0}}\lsb\frac{1}{\veps}-
  \gamma-\psi\lb\frac{1}{2}\rb-\frac{1}{\veps}-\ln y+\mathcal{O}(\veps)+\mathcal{O}(y^{\veps+1})+\ln y\rsb=\dots
\end{equation}
where \(\gamma\) is the Euler constant and \(\psi\) is the digamma function. The cancelation
of all dangerous divergencies is clearly seen. The final result of integration is
\begin{equation}
  \dots=-\frac{1}{4}\lb\gamma+\psi\lb\frac{1}{2}\rb\rb=-\frac{1}{4}(-2\ln2),
\end{equation}
where we have used the known value for \(\psi(1/2)\).

The third term in~\eqref{El} has been already integrated in the calculation of the small \(\lambda\)
asymptotics of the distance. This term can be rewritten in terms of incomplete beta
functions using~\eqref{beta},
\begin{multline}
\label{3rd_frac_betas}
  B_x\lb\frac{1}{4},-\frac{3}{2}\rb+2B_x\lb\frac{3}{4},-\frac{3}{2}\rb-
  2B_x\lb1,-\frac{3}{2}\rb-2B_x\lb\frac{5}{4},-\frac{3}{2}\rb+\\+
  B_x\lb\frac{7}{4},-\frac{3}{2}\rb-4B_x\lb2,-\frac{3}{2}\rb+
  4B_x\lb\frac{9}{4},-\frac{3}{2}\rb.
\end{multline}
The expansion of the first term in~\eqref{3rd_frac_betas} is
\begin{equation}
  B_x\lb\frac{1}{4},-\frac{3}{2}\rb\underset{x\to1}{=}B\lb\frac{1}{4},-\frac{3}{2}\rb+
  \frac{2}{3}(1-x)^{-3/2}+\frac{5}{3}(1-x)^{-1/2}+\mathcal{O}\lb(x-1)^{1/2}\rb.
\end{equation}
The expansion of other terms in~\eqref{3rd_frac_betas} is given in~\eqref{beta_exps}.
Taking into account the expansions~\eqref{beta_exps} and the coefficients in~\eqref{3rd_frac_betas}
we can check that all divergences are canceled, schematically
\begin{equation}
  \begin{aligned}
    1+2-2-2+1-4+4=0,\\
    \frac{5}{3}+2-2\cdot\frac{2}{3}-2\cdot\frac{1}{3}-\frac{1}{3}-4\cdot\frac{-2}{3}-4=0.
  \end{aligned}
\end{equation}

Collecting all the terms together, the expansion of energy at small \(\lambda\) becomes
\begin{multline}
  E=\frac{R^2}{4\pi\alpha'}\frac{c}{\lambda}\lsb B\lb-\frac{1}{4},\frac{1}{2}\rb+
  \frac{4\lambda}{3}\lb2\ln2+B\lb\frac{3}{4},-\frac{1}{2}\rb-B\lb1,-\frac{1}{2}\rb\rb+\right.\\\left.+
  \frac{8\lambda^2}{9}\lb B\lb\frac{1}{4},-\frac{3}{2}\rb+2B\lb\frac{3}{4},-\frac{3}{2}\rb-
  2B\lb1,-\frac{3}{2}\rb-2B\lb\frac{5}{4},-\frac{3}{2}\rb+\right.\right.\\\left.\left.+
  B\lb\frac{7}{4},-\frac{3}{2}\rb-4B\lb2,-\frac{3}{2}\rb+
  4B\lb\frac{9}{4},-\frac{3}{2}\rb\rb\rsb.
\end{multline}
Substituting the values of corresponding beta functions from~\eqref{beta_values1}
and from
\begin{equation}
\label{beta_values2}
  B\lb-\frac{1}{4},\frac{1}{2}\rb=-\frac{2}{\rho},\quad
  B\lb\frac{1}{4},-\frac{3}{2}\rb=\frac{5\pi\rho}{6},
\end{equation}
we obtain finally
\begin{equation}
\label{en_small_l}
  E\underset{\lambda\to0}{=}-\frac{R^2}{2\pi\alpha'\rho}\frac{c}{\lambda}\lsb1+
  \frac{2\lambda}{3}\lb1-2\rho(\ln2+1)\rb+
  \frac{4\lambda^2}{9}\lb\frac{1}{2}+4\rho-\frac{9\pi\rho^2}{2}\rb\rsb\!\!.
\end{equation}


The last step is to derive \(E(r)\) at small \(r\) using the expansions~\eqref{dist_small_l}
and~\eqref{en_small_l}. A straightforward way consists in making series reversion in~\eqref{dist_small_l}
and substituting \(\lambda(r)\) into~\eqref{en_small_l}.
Writing the expansion~\eqref{dist_small_l} as
\begin{equation}
  r\underset{\lambda\to 0}{=}a_1\lambda+a_2\lambda^2+a_3\lambda^3+\mathcal{O}(\lambda^4),
\end{equation}
where the series coefficients are equal to
\begin{equation}
  a_1=\frac{1}{\rho c},\quad
  a_2=\frac{2(2\rho-1)}{3\rho c},\quad
  a_3=\frac{4}{9\rho c}\lb\frac{1}{2}-4\rho+\frac{3\pi\rho^2}{2}\rb,
\end{equation}
we can obtain the reversed series \(\lambda(r)\),
\begin{equation}
\label{l_r_exp}
  \lambda\underset{r\to 0}{=}A_1r+A_2r^2+A_3r^3+\mathcal{O}(r^4),
\end{equation}
with coefficients
\begin{equation}
  \begin{aligned}
    &A_1\equiv a_1^{-1}=\rho c,\quad
    A_2\equiv -a_1^{-3}a_2=\frac{2\rho^2 c^2(1-2\rho)}{3},\\
    &A_3\equiv a_1^{-5}(2a_2^2-a_1a_3)=\frac{4\rho^3c^3}{9}\lsb\lb8-\frac{3\pi}{2}\rb\rho^2-4\rho+\frac{3}{2}\rsb.
  \end{aligned}
\end{equation}
From~\eqref{l_r_exp} we get also the series expansion for \(1/\lambda\),
\begin{equation}
\label{inv_l_r_exp}
  \frac{1}{\lambda}\underset{r\to 0}{=}\frac{1}{A_1r}-\frac{A_2}{A_1^2}+\lb\frac{A_2^2}{A_1^3}-\frac{A_3}{A_1^2}\rb r+\mathcal{O}(r^2).
\end{equation}
The substitution of~\eqref{l_r_exp} and~\eqref{inv_l_r_exp} into~\eqref{en_small_l} yields
\begin{equation}
\label{El2}
  E\underset{r\to0}{=}-\frac{R^2c}{2\pi\alpha'\rho}
  \lsb\frac{1}{A_1r}-\frac{A_2}{A_1^2}+\frac{B_1}{3}+
  \lb\frac{A_2^2}{A_1^3}-\frac{A_3}{A_1^2}+\frac{A_1B_2}{9}\rb r+\mathcal{O}(r^2)\rsb,
\end{equation}
where for compactness we temporarily introduced two new notations,
\begin{equation}
  B_1= 2\lb1-2\rho(\ln2+1)\rb,\quad
  B_2= 4\lb\frac{1}{2}+4\rho-\frac{9\pi\rho^2}{2}\rb.
\end{equation}
Inserting the values of the coefficients \(A_i\) and \(B_i\) into~\eqref{El2}
we find finally the ultraviolet asymptotics of potential energy,
\begin{equation}
\label{small_r_not_s2}
  E\underset{r\to0}{=}\frac{R^2}{\alpha'}
  \lsb-\frac{1}{2\pi\rho^2}\frac{1}{r}+\frac{2c\ln2}{3\pi}+
  \frac{2c^2\rho}{9\pi}\lb\lb3\pi+4\rb\rho-4\rb r+\mathcal{O}(r^2)\rsb.
\end{equation}

It is interesting to compare~\eqref{small_r_not_s2} with~\eqref{small_r_not_s},~\eqref{rho}.
The leading Coulomb contributions are equal, as it must be since the both models
share the same leading ultraviolet background determined by the metric of AdS space.
But the next-to-leading contributions are different: {\it It is a constant in the case of SW
model with linear dilaton and it is linear in $r$ in the SW model with quadratic dilaton}.
The linear in $r$ correction in~\eqref{small_r_not_s2} becomes the next-to-next-to-leading
contribution. Our calculation of short-distance energy~\eqref{small_r_not_s2}
turned out to be substantially more cumbersome than the calculation of~\eqref{small_r_not_s}
in~\cite{Afonin:2022aqt} because we needed to calculate this last contribution.

This remark is rather important because many discussions of holographic confinement potential
in the literature make use of a general relation for renormalized energy from the original analysis
of Ref.~\cite{Kinar:1998vq}. But that relation was derived using {\it one subtraction} of infinite constant
while we needed {\it two subtractions} and this caused the aforementioned subtleties in calculation of
renormalized energy. It is precisely because of these subtleties that we had to re-calculate the
energy from the very beginning. One can show that this situation occurs for any metric
deformation $h(z)$ in~\eqref{az_metric} of the form
\begin{equation}
\label{h3}
  h=e^{(kz)^n},
\end{equation}
when $0<n\leq1$, a discussion of this point will be presented elsewhere.
In particular, a study of the confining potential for the warp factor~\eqref{h3}
was performed in Ref.~\cite{heavy7}. One might conclude that our analysis is a particular case
of the analysis in~\cite{heavy7} for $n=1$. However, as we have emphasized above, this is not the case.

\subsection{Asymptotics of energy at large distances}

The large distances in the energy~\eqref{en_int} correspond to large \(\lambda\).
But the reality condition~\eqref{sqrt_exp} restricts the maximum value of \(\lambda\)
by \(\lambda=3/2\), see~\eqref{rc}, hence, the large distance behavior
of the potential energy should be derived from the asymptotic expansion at \(\lambda\to3/2\).

In our further analysis, we closely follow Ref.~\cite{Afonin:2022aqt} where we derived
the given infrared asymptotics in the scalar SW model with quadratic dilaton.
The main contribution to the integrals for the distance~\eqref{rl} and the energy~\eqref{unreg_en}
comes from the upper integration bound, \(v=1\), since the
integrals diverge in the upper limit. We need to expand the integrands around that
point. The corresponding expansion of expression under the square root is
\begin{equation}
  1-v^4e^{8\lambda(1-v)/3}\underset{v\to1}{=} A(\lambda)(1-v)+B(\lambda)(v-1)^2,
\end{equation}
\begin{equation}
  A(\lambda)= 4-\frac{8\lambda}{3},\qquad
  B(\lambda)=\frac{32\lambda}{3}-\frac{32\lambda^2}{9}-6,
\end{equation}
The other factors under the integrals either do not diverge or do not depend on
\(v\). This leads to the following expressions for the distance and energy
\begin{equation}
\label{r_v_1}
  r\underset{v\to1}{=}\frac{2\lambda}{c}\int\displaylimits_{0}^{1}
  \frac{dv}{\sqrt{A(\lambda)(1-v)+B(\lambda)(v-1)^2}},
\end{equation}
\begin{equation}
\label{en_v_1}
  E\underset{v\to1}{=}\frac{R^2}{\pi\alpha'}\frac{c}{\lambda}e^{4\lambda /3}\int\displaylimits_{0}^{1}\,
  \frac{dv}{\sqrt{A(\lambda)(1-v)+B(\lambda)(v-1)^2}}.
\end{equation}
Substituting the integral in~\eqref{r_v_1} into~\eqref{en_v_1} and taking
the limit \(\lambda\to3/2\) we get
\begin{equation}
\label{large_r_not_s3}
  E\underset{r\to\infty}{=}\frac{R^2}{\alpha'}\frac{2e^2c^2}{9\pi}r.
\end{equation}

It can be easily checked that the obtained asymptotics~\eqref{large_r_not_s3} satisfies
the general relation~\eqref{large_r_not_s2} for the infrared behavior of energy in confining
string theories. Indeed, $G_{00}$ in~\eqref{h2} reaches its minimum value at $z_0=\frac{3}{2c}$,
we obtain then $G_{00}(z_0)=4e^2c^2R^2/9$ and the relation~\eqref{large_r_not_s2} reproduces~\eqref{large_r_not_s3}.

\section{Concluding remarks}

We briefly reviewed the calculation of potential energy arising between static sources
within the framework of Soft-Wall holographic approach to strong interactions. In the case of positive quadratic
dilaton background, this calculation is known to result in a
Cornell-like potential. We argued further that qualitatively the same static potential must arise for a broad class
of modified SW models with positive dilaton background which do not lead to Regge spectrum. As an example we considered
in detail a particular case of other type of the dilaton background, namely a scalar SW model with ``linear dilaton''.
The spectrum of this model is Hydrogen-like, i.e., it has no Regge behavior even closely. However, the static potential
predicted by this model is very similar. Below we demonstrate a one unexpected aspect of numerical similarity.

The derived large and small distance behavior of energy, given by~\eqref{large_r_not_s3} and~\eqref{small_r_not_s2},
can be compactly rewritten as
\begin{equation}
\label{large_r_not_s3b}
  E\underset{r\to\infty}{=}\frac{R^2}{\alpha'}\sigma_\infty r,\qquad \sigma_\infty=\frac{2e^2c^2}{9\pi},
\end{equation}
\begin{equation}
\label{small_r_not_s2b}
  E\underset{r\to0}{=}\frac{R^2}{\alpha'}\lsb-\frac{\kappa_0}{r}+C+\sigma_0 r\rsb,
\end{equation}
where
\begin{equation}
\label{rho2}
  \kappa_0=\frac{1}{2\pi\rho^2},\,\,\,
  C=\frac{2c\ln2}{3\pi},\quad
  \sigma_0=\frac{2c^2\rho}{9\pi}\lb\lb3\pi+4\rb\rho-4\rb,\quad
  \rho=\frac{\Gamma\!\!\lb\frac{1}{4}\rb^2}{(2\pi)^{3/2}}.
\end{equation}
If we measure the energy in units of $R^2/\alpha'$, the quantity $\sigma$ becomes the slope of
linear potential. This slope is different at large and short distances.
In the Cornell potential~\eqref{cornell}, however, the slope is the same at all distances.
Consider the ratio of slopes in the large and small distance asymptotics derived for our
scalar SW model with linear dilaton,
\begin{equation}
\label{ratio1}
\text{Linear dilaton}:\qquad \frac{\sigma_\infty}{\sigma_0} = \frac{e^2}{\rho\lb\lb3\pi+4\rb\rho-4\rb}\simeq 1.23.
\end{equation}
It is seen that the difference in the slopes turns out to be not very significant,
especially taking into account that the bottom-up holographic approach is supposed
to describe QCD in the large-$N_c$ limit --- the discrepancy is roughly within the
accuracy of this limit.

Let us compare the ratio~\eqref{ratio1} with the analogous ratio in the scalar SW model with quadratic dilaton
which follows from~\eqref{large_r_not_s} and~\eqref{rho},
\begin{equation}
\label{ratio2}
\text{Quadratic dilaton}:\qquad \frac{\sigma_\infty}{\sigma_0} = \frac{e}{\pi\rho^2}\simeq 1.24.
\end{equation}
We see that the ratios in~\eqref{ratio1} and~\eqref{ratio2} are impressively close.
Note also that this ratio in the vector SW model with quadratic dilaton
coincides with~\eqref{ratio2} identically~\cite{Andreev:2006ct}.
All this indicates that within the SW bottom-up holographic approach the given ratio seems to be
approximately model independent.

\section*{Acknowledgements}

This research was funded by the Russian Science Foundation grant number 21-12-00020.

\end{document}